\documentclass[conference]{IEEEtran}
\IEEEoverridecommandlockouts
\usepackage{cite}
\usepackage{amsmath,amssymb,amsfonts}
\usepackage{graphicx}
\usepackage{textcomp}
\usepackage{xcolor}
\usepackage{makecell}
\usepackage{algorithm}
\usepackage[noend]{algpseudocode}
\def\BibTeX{{\rm B\kern-.05em{\sc i\kern-.025em b}\kern-.08em
    T\kern-.1667em\lower.7ex\hbox{E}\kern-.125emX}}
	
\providecommand{\tabularnewline}{\\}

\makeatletter
\def\BState{\State\hskip-\ALG@thistlm}
\makeatother

\begin{document}

\title{PlayNPort: A Portable Wireless Music Player and Text Reader System}

\author{\IEEEauthorblockN{Lakhan Shiva Kamireddy\IEEEauthorrefmark{1}, Dharmik Thakkar\IEEEauthorrefmark{1}, Lakhan Saiteja K\IEEEauthorrefmark{2}}
\IEEEauthorblockA{\IEEEauthorrefmark{1}VLSI and Embedded Systems Group, Department of Electrical and Computer Engineering, University of Colorado\\ Boulder, CO 80303, USA, Email: \{lakhan.kamireddy, dharmik.thakkar\}@colorado.edu}
\IEEEauthorblockA{\IEEEauthorrefmark{2} Department of Electronics and Electrical Communication Engineering,
Indian Institute of Technology Kharagpur\\ West Bengal 721302, India, Email: lakhansaiteja@gmail.com}
}

\maketitle

\begin{abstract}
Portable Consumer Electronics has made a mark in the industry. With the ease of use at an accessible price range, they have experienced significant growth in the market. Our idea is to develop a portable wireless music player and text reader using a Cortex-M series microcontroller and bare-metal programming techniques. We chose to use an SD card as the storage device. The resulting electronic device is similar to a consumer grade music player available in a car. The system comprises an MCU, an MP3 encoder/decoder, an LCD, an audio output jack, an SD card and a remote control. We also present various challenges involved in developing the system and solutions we used to overcome the challenges. The intricacy of the work lies in the fact that the system was developed to be consumer-centric by providing a rich User Experience. It can be used as a personal entertainment system in a car.
\end{abstract}

\section{Introduction}


With a plethora of embedded devices, the ubiquitous nature of these systems is felt across the consumer electronics industry.
One such portable electronic device is a music player. A variety of music players are commercially available. These are usually developed by several teams, each developing a set of features. Then they are assembled, packaged and marketed to the consumer as a product in the market. However, we are curious to learn what it takes to build a consumer grade electronic device from scratch, by putting ourselves into the shoes of a system hardware designer, a firmware engineer, and an algorithm designer.

The idea of this paper is to present the development of a complete embedded system in the lab, and understand the processes, development methodologies involved and challenges in implementing a complete system. We developed PlayNPort, a portable wireless music player, and text reader. It gives complete control of the system to the user through a custom remote. Its features comprise play/pause, play next/previous song, read a text, exit to menu. We implemented the system on a printed circuit board (PCB) and made use of off the shelf electronic components, a TI MSP432, decoder ICs and an SD Card from a retail store. In this paper, we explained the system design and implementation methodology.

The paper is organized into the following sections: Section II introduces the Board Design. In Section-III, we introduce the communication subsystem implementation Section-IV presents Firmware design. Challenges faced in development are presented in Section-V. Conclusion and Future work is presented in Section-VI.

\section{Board Design}
Fig 1. presents a block-level view of the embedded device. The system's brain is an ARM Cortex M4F microcontroller.  Following sections present board level design \& components involved.\\

\begin{figure}[!t]
\centering
\includegraphics[scale=0.5]{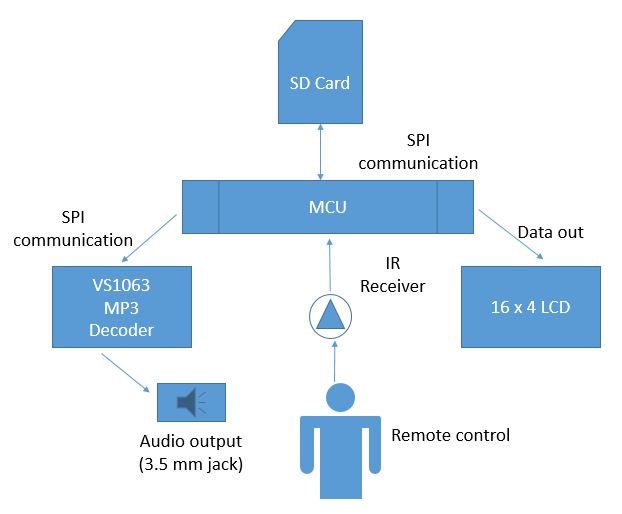}
\caption{Block Diagram of the Embedded System}
\end{figure}

\subsection{LCD Interface}
We have used a character LCD [1] to display the menu and messages on the screen. The character LCD only supports a 5V/0V logic level. The MSP432 General Purpose IO (GPIO) pins are capable of producing 3.3V/0V logic levels. We used bidirectional voltage level converters to convert the voltages back and forth.\\

\subsection{SD Card Interface}
We used an SD card breakout board. The SD card can be operated either in SD mode or SPI mode. For us, SPI mode was more suitable as we wanted more control over the card. An SD card has an inbuilt SD controller with an array of NAND flashes. We connected the Master-In Slave-Out (MISO), Master-Out Slave-In (MOSI) and Serial Clock (SCLK) pins to port pins 3.5, 3.6, 3.7 on the MSP432 that correspond to eUSCI B2 module [2].\\

\subsection{IR Module}
We used an IR module to give a user the control of the embedded system wirelessly using a custom IR remote that uses a lithium coin cell battery. Internally, it makes use of a microcontroller IC, which is a 4-bit RISC microcontroller with a built-in IR LED drive pin.\\
Remote control was used for IR transmission using an IR emitter LED on the head of the remote [3]. The remote has 9 buttons, each corresponding to a specific command. These include power, A, B, C, up, left, right, down and center buttons. These buttons are enough to give user controls like music play/pause, previous/next song, volume up/down and exit to menu options.\\

\subsection{MP3 decoder and 3.5 mm jack}
We chose to use the VS1063 MP3 decoder/encoder and codec to decode the mp3/wav format files stored on the SD card. The VS1063 consists of an inbuilt low-power DSP core, ROM memories, 16 kb instruction RAM, and 80 kb data RAM [4]. We used the SPI module and the common voltage buffer GBUF.
The TRRS 3.5 mm jack breakout board contains Tip, Ring 1, Ring 2, Sleeve pins. All the TRRS pins were connected to the respective contacts on the male jack.

\section{Communication Subsystem Implementation}

\subsection{Serial Peripheral Interconnect (SPI)}
SPI is used to establish communication between devices. SPI runs in a master/slave setup. It's capable of being run in full duplex mode [5]. When compared to Inter IC (I2C) Communication, SPI attains better speeds, often capable of reaching 20 Mbps. We configured the TI MSP432 as the master, SD Card and VS1063 IC as slaves. We used Enhanced Universal Serial Communication Interface (eUSCI) module on the MSP432 for data transfer. In synchronous mode, eUSCI can connect the device to an external system through either three or four pins. We operate in 3-pin mode to keep matters simple. SPI mode is selected when the USYNC bit is set and is controlled using UCMODEx bits [6].

We have an option to choose LSB-first or MSB-first data transmit and receive. In SPI, data shift and data latch are done on opposite clock edges. Hence, there are 4 different operating modes. Mode 0 (Clock Phase (CKPH) = 0, Clock Polarity (CKPL) = 0) means latch on a positive edge, then shift on the next negative edge. Mode 1 (CKPH = 0, CKPL = 1) means shift on positive edge, then latch on the next positive edge. Mode 2 (CKPH = 1, CKPL = 0) means shift on negative edge, then latch on the next positive edge. Please refer to the figure in [6] for SPI modes in MSP432.

The choice of mode of operation must be made appropriately by referring to the datasheet of the slave device (SD card and VS1063 IC) otherwise it may lead to data corruption. Along with this, bit rate and LSB (or MSB) first choices should also be made appropriately.

SPI mode can be controlled by four sets of registers [5].
\begin{enumerate}
\item General controls and bit clock generation registers
\item SPI data transmit control registers.
\item SPI data receiving control registers.
\item SPI interrupt related control registers.
\end{enumerate}
In the following section, we discuss some important configuration registers.

eUSCI Bx control word 0 register, UCBxCTLW0 is a general control register which is 16-bit wide [7]. Its bits can be set accordingly to configure settings like CKPH, CKPL, MSB-first, transfer data character length, master mode select, 3 or 4 pin mode select.

eUSCI Bx Bit rate control word register, UCBxBRW is a bit rate clock generation register which is 16-bit wide [7]. Its bits can be set accordingly to configure bit rate. UCBRx stores a pre-scaler value to divide the selected CKSE BRCLK to calculate the desired frequency of SPI clock UCLK. Equation 1. presents the formula used in the calculation. 

\begin{equation}
f_{uclk}=f_{BitClock} = f_{BRCLK}/UCBRx
\end{equation}

This is the bit transmit-receive rate.

The operational principle of eUSCI Bx module is simple. When data is moved to UCBxTXBUF register, known as the transmit data buffer, eUSCI module initiates the data transfer. The data in the UCBxTXBUF register is then moved to the TXS register when it is empty, initiating the data transfer as configured by the UCMSB setting [6]. The data on the UCBxSOMI line is shifted into the RXS register on the opposite clock edge.

When a complete character is received, the received data is moved from RXS register to the UCBxRXBUF. Data has to be read from UCBxRXBUF when the UCBRXIFG flag is set. A set on the transmit interrupt flag UCTXIFG only indicates that data has been moved from the UCBxTXBUF register to the TXS register and the UCBxTXBUF register is ready for new data [6]. It does not indicate a complete RX/TX transaction. To receive data into the eUSCI in master mode, dummy data must be written into the UCBxTXBUf register (in our case 0xFFh) because both transmit and receive operations occur simultaneously.\\
\subsection{NEC Protocol for Wireless IR Communication}
The NEC is a standard protocol using pulse distance encoding of the bits. Each pulse is 560μs long. Logic bit 1 takes 2.25ms to transmit and logical 0 takes 1.125ms [8].

A message starts with an Automatic Gain Control (AGC) pulse. A command is transmitted only once. A repeat code is transmitted every 110ms for as long as the key remains pressed. This repeat code is a 9ms high-speed AGC burst followed by a 2.25ms void space and a 560μs pulse [8].

\section{Firmware Design}
Firmware for this project comprises drivers for each of the new hardware interfaces (SD card, Audio Codec, IR decoder) and porting the existing drivers (character LCD drivers) for 8051 [10] onto MSP432 and developing application logic for playing the songs, music control (volume control, play/pause, previous/next songs) and LCD text reader application. The following sections outline the firmware we developed for PlayNPort.\\
\subsection{SPI Driver}
These drivers involve writing and reading registers of eUSCI\_B module. To save development time we coded the functions using polling technique which checks the corresponding interrupt flag for successful completion of data transfer. These can be made interrupt based by adding a circular buffer to prevent data corruption. Interrupt-based techniques improve efficiency while enabling CPU to sleep in low energy modes. A point to note here is that the driver to receive bytes makes use of the transmit driver. In SPI, the master has to send 0xFF as (dummy) data to generate a clock signal for the slave to transmit data.\\
\smallskip{}

\smallskip{}
\smallskip{}
\hfill{}TABLE I. SPI Driver functions\hfill{}

\smallskip{}

\smallskip{}
\begin{center}
\hfill{}%
\begin{tabular}{|c|c|}
\hline 
\makecell[l]{Function} & \makecell[l]{Description}\tabularnewline
\hline 
\hline 
\makecell[l]{spi\_init} & \makecell[l]{Initializes SPI module by defining clock\\ source, bitrate, MSB/LSB first mode,\\ master/slave mode, synchronous mode.}\tabularnewline
\hline 
\makecell[l]{Transmit} & \makecell[l]{Transmits data}\tabularnewline
\hline 
\makecell[l]{rcvr\_spi} & \makecell[l]{Receives a byte}\tabularnewline
\hline 
\makecell[l]{rcvr\_spi\_multi} & \makecell[l]{Receives multiple bytes.}\tabularnewline
\hline 
\end{tabular}
\hfill{}
\end{center}

\subsection{SDHC drivers}
SD Card is a Non-Volatile Memory (NVM). SD cards respond to a set of predefined commands, that are specified by the SD card association. Six registers are defined within the card interface namely Operating Conditions Register (OCR), Card identification register (CID), Card specific data (CSD), RCA, Driver state register (DSR), SD card configuration register (SCR) [9]. These can be accessed only by corresponding commands.\\
We configured the SD card to be accessed in SPI mode [14]. All data tokens to the SD card are 8-bit and always byte aligned to the CS signal. In our SPI drivers, we implemented methods to receive SD card response (receive\_response), wait for the card to be ready (card\_ready), select and deselect the card (select, deselect), configure GPIO and SPI modules (power\_on), receive a data packet from SD card (rcvr\_datablock), send a specific command to SDC (send\_cmd), send reset command and initialize SD in SPI mode (disk\_initialize), get SDC status (disk\_status), read SDC sectors (disk\_read), device timer function (disk\_timeproc).

The following are steps to successfully initialize the SDHC [9].
\begin{enumerate}
\item CMD0, argument: 0x0, CRC: 0x95 and the response we got was 0x01.
\item CMD8, argument: 0x000001AA, CRC: 0x87 and the response we got was 0x01.
\item a) CMD55, argument: 0x0, CRC: 0x65, and the response we got was 0x01.
      b) ACMD41, argument: 0x40000000, CRC: 0x78, if the response if 0x0 we are OK, if it is 0x01 we go to step 3.
\end{enumerate}

After performing an SD card sector write, we get one of the following responses [10].
\begin{enumerate}
\item 010 $->$ Data accepted
\item 101 $->$ Data rejected due to CRC error
\item 110 $->$ Data rejected due to write error and bit 0 is a one
\end{enumerate}

\subsection{FAT file system drivers}
A filesystem is the collection of many methods and data structures that an operating system needs to exercise control over the storage and retrieval of data. The file system used on the MMC/SD Card is FAT. The MMC/SDC specifications define the FAT type as FAT32 for 4GB to 32GB. As we use a 16 GB SD Card High Capacity (SDHC), the FAT type is FAT32. Only a FAT volume can co-exist on the card along with an FDISK partition.

We have ported the generic FAT file system [11], FatFs [10] into our application and implemented the MSP432 platform-specific methods. The resultant file system drivers are used to control the SDHC. We implemented the following methods. The method f\_mount is used to register/unregister the work area of the volume. f\_open method is used to create/open a file. f\_close is used to close a file. f\_read is a method used to read data.

\begin{figure}[!t]
\centering
\includegraphics[scale=0.5]{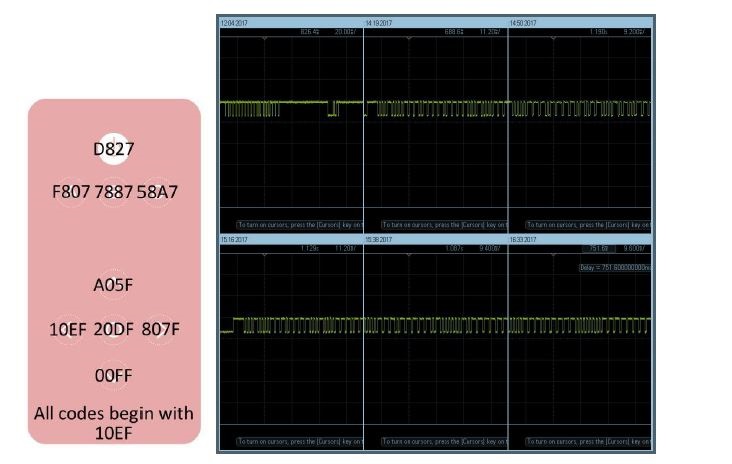}
\caption{A. Codes B. IR signals received for A, B, C, $<, \wedge, >$ button press}
\end{figure}

\subsection{Audio decoder drivers}
To enable VS1063 IC [12] for audio decoding operation, we wrote certain drivers making use of the SPI methods we implemented. Table 2. presents the functions and their descriptions below.\\
\smallskip{}

\smallskip{}
\smallskip{}
\hfill{}TABLE II. Audio Decoder Driver functions\hfill{}
\smallskip{}
\begin{center}
\hfill{}%
\begin{tabular}{|c|c|}
\hline 
\makecell[l]{Function} & \makecell[l]{Description}\tabularnewline
\hline
\hline 
\makecell[l]{Spiinit} & \makecell[l]{Initializes SPI and GPIO modules.\\ 
It sets initial SPI \\clock frequency as 2MHz \\with MSB first and Mode 2.}\tabularnewline
\hline 
\makecell[l]{WriteSpiByteSDI} & \makecell[l]{Writes Serial Data Interface \\(SDI) data of the codec.}\tabularnewline
\hline 
\makecell[l]{WriteSdi} & \makecell[l]{Writes multiple \\Serial Data Interface \\(SDI) bytes. Returns 0 on success. \\Returns -1 if the number of \\bytes is greater than 32}\tabularnewline
\hline 
\makecell[l]{ReadSci} & \makecell[l]{Reads Serial Command Interface \\(SCI) registers.}\tabularnewline
\hline 
\makecell[l]{WriteSci} & \makecell[l]{Writes Serial Command Interface \\(SCI) registers.}\tabularnewline
\hline 
\makecell[l]{VSTestInitHardware} & \makecell[l]{Hardware Initialization for \\VS1063. Configures CS, Reset and \\DREQ pins.}\tabularnewline
\hline
\makecell[l]{VSTestInitSoftware} & \makecell[l]{Software Initialization for VS1063. \\Configures SMOD register and\\ chip select for command \\and data interfaces by enabling \\SM\_SDISHARE, performs \\a quick sanity check by \\writing two registers and \\reading them for validity check, \\sets clock frequency to \\a higher clock, sets volume, \\sets WRAM address and \\finally loads latest \\VS1063 Patches package.}\tabularnewline
\hline
\end{tabular}
\hfill{}
\end{center}
\subsection{IR Transmission, Reception and Decoding}
The NEC IR transmission protocol uses pulse length encoding of the message bits.
When a key is pressed on the remote control, the message transmitted consists of the following.
A start bit, which is recognized by a 9ms leading pulse burst. Then follows a 4.5 ms space. Then follows 8-bit address for the receiving device. Then follows 8-bit logical inverse of the address. This follows 8-bit command. Then follows 8-bit logical inverse of the command. Then follows a 562.5 μs pulse burst to signify the end of the message transmission [8].
As we see, 32 bits have to be read after the start bit is received.\\

\subsection{IR Decoding Logic}
We used the MSP432 timer for decoding the IR codes. The timer is configured to make use of the SMCLK clock source and is configured to run in continuous up counting mode, whenever needed. Otherwise, it is halted temporarily, and its counts are reset to zero.
A falling edge is initially detected using an interrupt. An interrupt is generated whenever a falling edge is detected on a port pin. Whenever the interrupt service routine is executed, the timer starts counting. If a start condition is detected, then we execute the program for storing the timer values in an array.
The algorithm is described as follows.
After every bit read, we halt the timer, clear the timer register values, start the timer in continuous mode.
\begin{algorithm}
\caption{Store timer data algorithm}\label{euclid}
\begin{algorithmic}[1]
\Procedure{DataStore}{}
\State $\textit{DetectStartCond}$
\State $\textit{Wait} \text{ till end of start bit }$
\BState \emph{top}:
\For {$i = 0, i < \textit{32}$}
\State $\textit{HaltTimer}$
\State $\textit{Clear Time Register}$
\State $\textit{StartTimer} \text{ in continuous mode}$
\State $timer[i] \gets \textit{timerVal}$
\State $i++$
\EndFor
\State $\textit{RaiseDataRecvFlag}$
\EndProcedure
\end{algorithmic}
\end{algorithm}

We continuously poll for data received flag in our program, when the flag is raised, we decode the array to a command. The decode logic is presented in Algorithm 2 [15].
Button A: Starts playing music. Enters music mode. In this mode, the user has options to play/pause, play next/previous song, increase/decrease the volume.
Button B: Starts reading the text file. Enters reading mode. In this mode, the user has an option to scroll down by choosing down arrow button.
Button C: We exit the current mode and go back to the menu.

\begin{algorithm}
\caption{IR Decode algorithm}\label{euclid}
\begin{algorithmic}[1]
\Procedure{CharacterDecode}{}
\State $\textit{Poll} \text{ for data received flag}$
\State $\textit{Match each timer value to a bit}$
\BState \emph{DataRecvd}:
\If {$bit[0:16] ==  \textit{0x10EF}$}
\State $\textit{DecodeRemainingBits}$
\State $\textit{MatchToACommand}$
\Else
\State $\textit{UnrecognizedCommand}$
\EndIf
\State $\textit{ClearDataRecvFlag}$
\EndProcedure
\end{algorithmic}
\end{algorithm}

\subsection{LCD Driver}
We have written methods for LCD driver to initialize it, read data, poll LCD busy flag, put command/data on LCD's data lines, display a character/string, go to desired DDRAM/CGRAM address, print main menu, scroll through text file, clear screen, read the text file from SD card to be displayed on LCD, populate text/music files, display a custom character, display custom character for play, pause, next, previous song. Initially, some of these methods were written for 8051. Then we ported them to MSP432 platform by making appropriate code tweaks.

\section{Challenges}
We faced many challenges in making the system functional and consumer friendly. Some of the problems and our solutions to overcome these are presented in the following sections.

\subsection{Issue mounting FAT file system}
Initially, when we attempted to execute f\_mount(), it returned FR\_DISK\_ERR. After hours of debugging, we found that it was the following check that didn't pass.
$if (ld\_word(fs->win + BS\_55AA)!=0xAA55) return 3;$
This checked for last two bytes (byte 511 and 512) of sector 0 to be equal to 0xAA55. Upon researching, we found that it is the signature of a Master Boot Record (MBR). We realized that somehow our sector 0 was corrupted. The solution was that, instead of using quick format option while formatting the SD card, we perform a complete format. Then f\_mount returned FR\_OK.

\subsection{Software initialization of VS1063 Audio Decoder}
Software initialization of VS1063 [12][13] included a sanity check in which it writes some data to registers and reads the same data for verification. This check failed due to data corruption, and when we analyzed it using signal analyzers, we found that the clock polarity and clock phase settings were misconfigured. We reconfigured them, and it started working.

\subsection{VS1063 not responding to SDI data writes}
Ideally, as soon as one starts writing valid data to the SDI, SDI decodes it to generate audio output. However in our case, all we could hear was some noise. Eventually, we found out that chip select for sending data is very different from chip select for sending commands. We fixed it by modifying firmware appropriately.

\subsection{Music plays in fast mode}
Initially, the SPI clock frequency is set to 2MHz for initialization. This led to the misbehavior. After initialization, we reduced the clock frequency to 120kHz, and it started to play just fine.

\section{Conclusion}
In this paper, we have developed PlayNPort, a portable wireless music player, and text reader. We made use of off the shelf electronic components, MSP432, decoder ICs and a PCB. We relied upon various datasheets, application notes and community resources in the process. By writing bare-metal firmware code, we achieved full control over the system. Thus, we were able to provide a rich user experience, by giving a user the options to play/pause, play next/previous song, scroll down while reading a text document. To improve the system, we can use a graphical LCD instead of a character LCD. As we tasked ourselves with building a complete system under budget, we chose a minimal configuration system. The total cost of the system is under 70\$.

As we know that IR is not the latest technology, in future, the system could be built using Bluetooth low energy (BLE) modules. IR needs a direct line of sight with the sensor. By going for Bluetooth we eliminate this need. We could also explore other VCD options that run at 3.3V, so we eliminate the use of voltage converters. We could go for an LCD that supports SPI communication to save many pins on the microcontroller that we consumed for data lines.

\section{Acknowledgement}
We would like to thank Prof. Linden McClure for providing us with the opportunity to work in the Embedded Systems Design lab at the University of Colorado Boulder. We would also like to thank our colleagues for being supportive through our endeavor. We would also like to thank the ECE department at CU Boulder.


\begin{thebibliography}{00}
\bibitem{b1} \emph{Dot Matrix Liquid Crystal Display Controller Datasheet}, Hitachi, pp 167-226, 2010.
\bibitem{b2} \emph{Using MSP432 serial modules}, Online, 2015, Accessed from http://www.samlewis.me/2015/05/using-msp432-eUSCI/. 
\bibitem{b3} \emph{IR Receiver Modules for Remote Control Systems}, Vishay Semiconductors, pp 1-7, 2008.
\bibitem{b4} \emph{VS1063a Encoder and Audio Coded circuit}, VLSI Solution, pp 1-93, v 1.31, 2017.
\bibitem{b5} \emph{Serial Peripheral Interface (SPI) User Guide}, Texas Instruments, pp 1-51, 2012.
\bibitem{b6} Bai~Ying, \emph{Microcontroller Engineering with MSP432: Fundamentals and Applications}, CRC Press, 2016.
\bibitem{b7} \emph{Enhanced Universal Serial Communication Interface (eUSCI) – SPI Mode}, SLAU424F, 2012, Texas Instruments.
\bibitem{b8} \emph{NEC IR Remote Control Interface}, Altium Techdocs, Online, 2017, Accessed from https://techdocs.altium.com/display/FPGA/NEC+Infrared+Transmission+Protocol.
\bibitem{b9} \emph{Sandisk microSDHC OEM Product Manual}, pp 1-25, Western Digital Inc., 2016.
\bibitem{b10} Elm,~Chang, \emph{How to use MMC/SSD}, 2010, Accessed from http://elm-chan.org/docs/mmc/mmc\_e.html.
\bibitem{b11} Elm,~Chang, \emph{FAT Filesystem}, 2010, Accessed from http://elm-chan.org/docs/fat\_e.html.
\bibitem{b12} \emph{VS1063 AppNote: Playback And Recording}, VLSI Solution, pp 1-7, v 1.10, 2016.
\bibitem{b13} \emph{VS1063a Patches: VLSI Solution Audio Decoder/Encoder}, VLSI Solution, pp 1-16, v 2.01, 2017.
\bibitem{b14} Stefan~Schauer, Christian~Speck, \emph{App Note: Interfacing the MSP430 With MMC/SD Flash Memory Cards}, Texas Instruments, pp 1-5, 2008.
\bibitem{b15} K~Lakhan~Shiva, 2017, \emph{Wireless Music Player and Text Reader using TI MSP432}, Accessed from https://github.com/lakhanshiva/SDCardController
\end{thebibliography}
\end{document}